\documentclass[aps,pra,epsf,twocolumn]{revtex4}
\usepackage{graphicx}
\usepackage{subfigure}
\usepackage{bm}

\usepackage[intlimits]{amsmath} 

\usepackage{verbatim} 
\usepackage{psfrag}

\usepackage{amssymb} 
 \newcommand{\bra}[1]{\langle
#1  | } \newcommand{\ket}[1]{ | #1 \rangle  }

\newcommand{\V}[1]{{\bf #1}}

\newcommand{\kin}{-\frac{\hbar^2 \nabla^2}{2m}}

\begin{document}

\title{Light scattering by ultracold atoms in an optical lattice}

\author{Stefan Rist,$^{1}$ Chiara Menotti,$^{2}$ and Giovanna Morigi$^{1,3}$}

\affiliation{
$^1$ Departament de F{\'i}sica, Universitat Aut\`onoma de Barcelona, 08193 Bellaterra, Spain\\
$^2$ CNR-INFM BEC, and Dipartimento di Fisica, Universit\`a di
Trento, I-38050 Povo, Italy\\
$^3$ Theoretische Physik, Universit\"at des Saarlandes, D-66041 Saarbr\"ucken, Germany}

\begin{abstract}

We investigate theoretically light scattering of photons by
ultracold atoms in an optical lattice in the linear regime. A full quantum theory for the atom-photon
interactions is developed as a function of the atomic state in the
lattice along the Mott-insulator -- superfluid phase
transition, and the photonic scattering cross section is evaluated
as a function of the energy and of the direction of emission. The
predictions of this theory are compared with the theoretical
results of a recent work on Bragg scattering in
time-of-flight measurements [A.M. Rey, {\it et al.}, Phys. Rev. A
{\bf 72}, 023407 (2005)]. We show that, when performing Bragg
spectroscopy with light scattering, the photon recoil gives rise
to an additional atomic site to site hopping, which can interfere
with ordinary tunneling of matter waves and can significantly
affect the photonic scattering cross section.
\end{abstract}

\date{\today}
\maketitle

\section{Introduction}

Bragg scattering in condensed matter is a powerful method for
gaining information over the structural properties of crystalline
solids. Usually, one employs thermal neutron beams, whose thermal
wavelength is of the order of the interparticle distance inside
the crystal. While elastic scattering allows one to measure the
reciprocal lattice primitive cell, inelastic scattering gives
information about the phonon spectrum and
anharmonicities~\cite{Ashcroft}. In atomic systems, Bragg
scattering has been applied for demonstrating long--range order in
structures of cold ions in traps~\cite{IonCrystals} and neutral
atoms in optical
lattices~\cite{Hemmerich,Phillips,Grynberg,Guidoni,Zimmermann_Bragg}.
Moreover it has proven to be a precise tool for the measurement of
the elementary excitations of trapped Bose-Einstein condensate
\cite{Ketterle,Davidson} and strongly-correlated atoms in optical
lattices \cite{Inguscio}. The spectra of the scattered photons,
moreover, provide information on the details of atom-photon
interactions. Studies on opto-mechanical systems, for instance,
showed that the Stokes and anti-Stokes components of the scattered
light may exhibit entanglement, which emerges from and is mediated
by the interaction with the quantum vibrational modes of the
scattering system~\cite{Pirandola03}. Such correlations are
endorsed by quantum interference in the processes leading to
photon scattering, which is mainly visible in the height of the
spectral peaks as a function of the emission angle~\cite{Cirac92},
and can be an important resource for quantum
networks~\cite{Mancini03,Morigi06}.

In this paper we investigate the opto-mechanical properties of
strongly-correlated atoms in optical lattices. These systems
present peculiar features, when compared with solid-state
crystals. In optical lattices the bulk periodicity is determined
by the light potentials, and is hence of the order of half the
laser wavelength. One remarkable property is that light both
couples to the atomic transition and is diffracted by the
crystalline structure which the atoms
form~\cite{Hemmerich,Phillips,Grynberg}. This property implies,
for instance, that the system may exhibit peculiar
self-organization, being the atoms a diffracting medium for the
light which traps
them~\cite{Phillips,Deutsch,Hemmerich_PRA,Hemmerich2}.

In the dispersive regime, when the optical lattice can be
considered a conservative potential, various states of ultracold
matter can be realized~\cite{Bloch-Review}, thereby mimicking
solid state models~\cite{Jaksch_BH,Maciej-Review}, a prominent
example of which is the quantum phase transition between a
Mott-insulator and a superfluid state~\cite{Greiner}. Bragg
spectroscopy provides an important tool for characterizing the
quantum state of the atomic gas~\cite{Ketterle,Davidson,Inguscio}.
The experimental procedure typically uses two laser beams, whose
wave vector difference ${\bf q}$ gives, by means of the mechanical
effects induced by photon recoil, a momentum and energy transfer
$\hbar {\bf q}$ and $\hbar\omega$~\cite{Ketterle}. The
corresponding atomic response is detected by a time-of-flight
measurement, consisting in releasing the trap and measuring the
momentum distribution by atom detection~\cite{Ketterle,Davidson}.
An alternative procedure makes use of parametric amplification
followed by time-of-flight measurement, thereby revealing the
energy transfer and the spectrum~\cite{Stoeferle,Tozzo}.
These procedures may allow one to measure the structure form
factor~\cite{Brunello,ReyBraggScattering,Menotti} and characterize
the state of the gas.

Most recently, ultracold atoms were loaded inside of optical
resonators, and first measurements of the spectrum of transmission
of the light at the cavity output showed novel features, which can
be brought back to the collective and coherent interaction of the
atoms with the
light~\cite{Vuletic,Hemmerich2,Esslinger,Reichel,Stamper-Kurn}.
Several theoretical works pointed out that the observation of the
photon scattered by ultracold atoms may provide complementary
information on the quantum state of the
atoms~\cite{Mekhov_Maschler,Meystre,Priscilla,RitschNature07,LarsonPRL},
which could be non-destructive in some
setups~\cite{Mekhov_QND,RitschNature07}.

\begin{figure}[htp] \centering
\includegraphics[width=.44\textwidth]{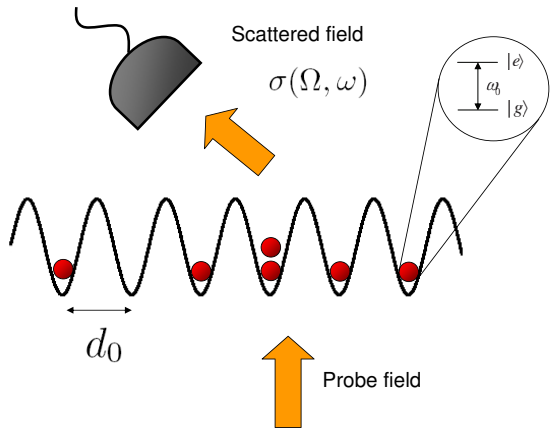}
\caption{\label{fig:setup}Light scattering by atoms trapped in a
one-dimensional optical lattice with lattice constant $d_0$. The
atoms are probed by a laser beam, with wave vector ${\bf k_L}$ and
frequency $\omega_L$, which couples to the atomic dipole
transition at frequency $\omega_0$ with ground state $|g\rangle$
and excited state $|e\rangle$ (see inset). The spectrum of the
scattered light is measured at a detector as a function of the
angle of emission. In experiments, one can also use a second laser beam, into which
the photon is emitted with high probability, hence implementing stimulated Bragg scattering~\cite{Ketterle}.}
\end{figure}

We also remark that theoretical studies on Fermionic systems in an optical lattice showed that intensity fluctuations of the scattered light may allow one to determine the temperature of the atomic cloud~\cite{Ruotekoski09}. Optical detection, and in particular the intensity of the Bragg peaks, were proposed as a mean for revealing fractional particle numbers of Fermi gases confined by optical lattices~\cite{Javanainen}.
In this paper, we study light scattering by ultracold bosonic atoms in an
optical lattice, in the setup sketched in Fig.~\ref{fig:setup}. We
use a full quantum description of the photonic and atomic fields,
for a range of optical lattice depths which covers the superfluid
to Mott-insulator transition. By starting from the general
Hamiltonian, we carry out the tight-binding and single-band
approximations, and we determine the scattering cross section of
photons in the linear response regime. Extending previous
works~\cite{Mekhov_Maschler}, we systematically take into account
the finite tunneling rate in evaluating the scattering cross
section for parameters sweeping along the phase
transition Mott-insulator to superfluid state. Our study focusses
on a small lattice of 7 sites, and solves numerically the
Bose-Hubbard model for this system. In order to get insight into
the numerical results, we also develop an analytical theory, which
extends the theory presented in~\cite{ReyBraggScattering} by
including the hopping induced by photon recoil.
The interference between the finite tunneling rate and
the photon-induced hopping is visible in the height of the Stokes
peaks as a function of the emission angle and can be revealed
experimentally.

This article is organized as follows: In Sec.~\ref{Sec:2} we
present the theoretical model. In Sec.~\ref{Sec:3} the scattering
cross section is evaluated both analytically and by means of
numerical simulations. The conclusions are discussed in
Sec.~\ref{Sec:4}.

\section{The Model} \label{Sec:2}

The scattering system we consider is composed by $N$ identical
bosonic atoms of mass $m$ in a periodic potential, as shown in
Fig.~\ref{fig:setup}. The relevant internal degrees of freedom of
the atoms are the electronic ground state $|g\rangle$ and an
excited state $|e\rangle$ that form a dipolar transition with
dipole moment ${\bf D}$ at the optical frequency $\omega_0$, which
couples to a weak laser probe. The Hamiltonian in second
quantization reads $H=H_{\rm at}+H_{\rm emf}+H_{\rm int}$
with~\cite{Maciej}

\begin{eqnarray}
&&H_{\rm at}=\hbar \omega_0\int {\rm d}{\bf r}\psi_e^{\dagger} ({\bf r})\psi_e({\bf r})+\sum_{j=e,g}H_j+H_{eg}\,,\\
&&H_{\rm emf}=\sum_{\lambda} \hbar \omega_{\lambda} a^{\dagger}_{\lambda} a_{\lambda}\,,
\end{eqnarray}
where by $\psi_j({\bf r})$ and $\psi_j^{\dagger}({\bf r})$ we
denoted the annihilation and creation operators of an atom in the
internal state $j=g,e$ at position ${\bf r}$, and by $a_{\lambda}$
and $a_\lambda^{\dagger}$ the annihilation and creation operators
of a photon in the mode at frequency $\omega_\lambda$, wave vector
${\bf k_\lambda}$ and polarization $\epsilon_\lambda\perp {\bf
k_\lambda}$. The atomic field operators obey the bosonic
commutation relations $[\psi_j({\bf r}),\psi_{j'}({\bf
r'})]=[\psi_j^{\dagger}({\bf r}) ,\psi_{j'}^{\dagger}({\bf r'})
]=0$ and $[\psi_j({\bf r}),\psi_{j'}^{\dagger}({\bf r'})
]=\delta_{jj'}\delta({\bf r}-{\bf r'})$. The Hamiltonian term
$H_g$, ($H_e$) describes the motion of the atoms in the internal
state $|g\rangle$ ($|e\rangle$), and $H_{eg}$ gives the
collisional interaction between the atoms in states $|g\rangle$
and $|e\rangle$. We will assume that the atoms interact with
radiation far-off resonance from the dipolar transition, hence the
occupation of the excited state is small and will be neglected.
Therefore, we just need to provide the detailed form of the ground
state term,

\begin{eqnarray}
H_g &=& \int {\rm d}{\bf r} \psi_g^{\dagger}({\bf r}) \left ( \kin
+V({\bf r}) \right ) \psi_g({\bf r}) \nonumber \\
& & +\frac{u_{gg}}{2} \int {\rm d}{\bf r} \psi_g^{\dagger}(\V{r})
\psi_g^{\dagger}(\V{r}) \psi_g(\V{r}) \psi_g(\V{r})\,, \label{H:g}
\end{eqnarray}
where $u_{gg}$ is the strength of the contact interaction. The
potential $V({\bf r})$ is assumed to be periodic along the
$x$-direction and reads

\begin{equation} \label{pot}
V({\bf r}) = V_0 \sin^2\left(\frac{\pi
x}{d_0}\right)+\frac{1}{2}m\omega_r (y^2+z^2)\,,
\end{equation}
where $V_0$ is the lattice depth, $d_0$ the lattice constant, and
$\omega_r$ the frequency of the harmonic trap which tightly confines
the transverse motion.

Finally, the interaction term between atoms and light reads (in
the length gauge)
\begin{equation}
H_{\rm int}=\sum_\lambda \hbar C_\lambda \int {\rm d}{\bf
r}\psi_e^{\dagger}({\bf r})
\psi_g({\bf r})a_\lambda {\rm e}^{{\rm
i}{\bf k_\lambda}\cdot {\bf r}}+{\rm H.c.}\,,
\end{equation}
where

\begin{eqnarray} \label{C:mu}
C_{\lambda} &=& \sqrt{\frac{\omega_{\lambda}}{2 \hbar \varepsilon_0
\mathcal V}} \left( \V{D}\cdot \V{\epsilon}_{\lambda} \right )
\end{eqnarray}
is the coupling strength, with $\varepsilon_0$ the vacuum electric
permittivity and $\mathcal V$ the quantization volume.

\subsection{Linear response}

At room temperature and equilibrium, the atoms are in the
electronic ground state and the state of the optical modes of the
electromagnetic field can be approximated with the vacuum
$|0\rangle$. We now assume that a laser, at frequency $\omega_L$
and wave vector ${\bf k_L}$, couples to the atomic dipole
transition. The laser field is described by a coherent state of
the corresponding electromagnetic field mode with amplitude
$\alpha_L$, such that the mean number of photons is given by
$|\alpha_L|^2$. In the regime in which the atom-laser coupling is
sufficiently weak, corresponding to the condition
$|C_L\alpha_L|\ll |\omega_0-\omega_L|$, we eliminate the excited
state from the equations of motion of the ground state in second-order perturbation theory in the small parameter $|C_L\alpha_L|/|\omega_0-\omega_L|$. The dynamics of the atoms in
the electronic ground state $|g\rangle$ is now described by the
effective Hamiltonian

\begin{equation}\label{H:tot}
H_{\rm eff}=H_g+H_{\rm emf}+H_{\rm int}^{\prime}\,,
\end{equation}
where the interaction term takes the form

\begin{eqnarray}\label{eqn:Heff}
H_{\rm int}^{\prime} &=&  \hbar \sum_{\lambda,\lambda'}
\frac{C_\lambda^* C_{\lambda'}}{\omega_{\lambda'}-\omega_0}
a_{\lambda}^{\dagger}a_{\lambda'} \int {\rm d}{\bf r} {\rm e}^{{\rm
i}{\bf q}\cdot {\bf r}}\psi_g^{\dagger}({\bf r})\psi_g({\bf r})\nonumber\\
 &=&  \hbar \sum_{\lambda,\lambda'}
\frac{C_\lambda^* C_{\lambda'}}{\omega_{\lambda'}-\omega_0}
a_{\lambda}^{\dagger}a_{\lambda'} \mathcal N_{\bf q}
\end{eqnarray}
and describes the absorption of a photon in the mode $\lambda'$
and wave vector ${\bf k_{\lambda'}}$and the emission into the mode
$\lambda$ and wave vector ${\bf k_{\lambda}}$, weighted by the
Fourier transform of the density $\mathcal N_{\bf q}=\int {\rm
d}{\bf r} {\rm e}^{{\rm i}{\bf q}\cdot {\bf
r}}\psi_g^{\dagger}({\bf r})\psi_g({\bf r})$, with
\begin{equation} {\bf q}={\bf k_{\lambda'}}-{\bf k_{\lambda}}\,.
\end{equation}
In the following we will assume that the interaction between
photons and atoms is essentially Hamiltonian, and hence fully
determined by the Schr\"odinger equation governed by
Eq.~(\ref{H:tot}). This is valid in the regime which we consider
in this article, namely, when the detuning of the light
$|\omega_0-\omega_L|\gg\gamma$, with $\gamma$ the linewidth of the
excited state.

In this work, we study Bragg scattering of laser photons by
atoms in the one-dimensional periodic array given by
potential~(\ref{pot}). We will hence evaluate the differential
scattering cross section for coherent scattering. Assuming that
$|\alpha_L|\ll 1$, so that the atoms absorb at most one photon from
the laser at a time, the differential scattering cross section is
found from the rate of scattering one laser photon
into the mode $\lambda$. In particular, the scattering rate reads

\begin{eqnarray} \label{rate}\Gamma_{\lambda_L\to
\lambda}&=&\frac{2\pi}{\hbar^2}\sum_{f}\left |
\bra{1_\lambda,f}H_{\rm int}'\ket{1_{\lambda_L},i}\right |^2\\
&\times&\delta^{(T)}(\omega_L-\omega_{\lambda}+(E_i-E_{f})/\hbar)\,,\nonumber
\end{eqnarray}
where we denoted by
$|1_{\lambda}\rangle=a^{\dagger}_{\lambda}|0\rangle$ the state of
the electromagnetic field with one photon in mode $\lambda$, and
by $|i\rangle$ and $|f\rangle$ the states of the atoms before and
after the scattering, respectively, which are eigenstates of
Hamiltonian $H_g$ at energies $E_i$ and $E_f$. The function

\begin{equation}
\delta^{(T)}(\omega)=\frac{\sin(\omega T/2)}{\pi \omega}
\end{equation}
is the diffraction function, giving energy conservation for
infinite interaction times,
$\lim_{T\to\infty}\delta^{(T)}(\omega)=\delta(\omega)$~\cite{Cohen}.

Equation~(\ref{rate}) shows clearly that the scattering rate
depends on the state of the atoms before and after the scattering
event. In the following, we derive the atom-light interaction
Hamiltonian in the tight-binding approximation and conclude this
section by introducing the many-body atomic states which are
relevant for the scattering process considered here.

\subsection{Tight-binding regime}

We assume that the atomic wavefunctions are well localized at the
lattice minima, such that the tight-binding approximation can be
applied. Furthermore, at ultralow temperature and not too strong
interactions, the atomic gas is in the lowest band of the periodic
potential and in the ground state of the radial oscillator, so
that the atomic field operator can be decomposed as

\begin{equation}
\label{eqn:wannier}
 \psi_g({\bf r})=
\phi_0(\rho) \sum_l w_l(x)  b_l\,,
\end{equation}
where  $w_l(x)=w(x-l d_0)$ is the Wannier function centered
at position $l d_0$, with the sum going over all lattice sites, and
$\phi_0(\rho)=\exp(-\rho^2/2\xi_r^2)/(\xi_r \sqrt{\pi})$ is the
ground state of the radial oscillator ($\rho=\sqrt{y^2+z^2}$) with
$\xi_r=\sqrt{\hbar/m \omega_r}$. The operators $ b_l$ annihilate
an atom at site $l$ and fulfill the standard bosonic commutation
relations $[ b_{l}, b^{\dagger}_{l'}]=\delta_{l,l'}$. Using this
decomposition in Eq.~(\ref{H:g}), allowing only nearest-neighbour
hopping and restricting to on-site atom-atom interactions, we
obtain the Bose-Hubbard Hamiltonian~\cite{Jaksch_BH}

\begin{eqnarray}
 H_g' =-J \sum_l   b_l^{\dagger} ( b_{l-1}+
b_{l+1}) +\frac{U}{2}\sum_l  n_l( n_l-1) -\mu \sum_l
n_l\,,\nonumber\\
\label{eqn:Hatomic}
\end{eqnarray}
where $ n_l= b_l^{\dagger} b_l $ is the atomic number operator at
site $l$ and $\mu$ is the chemical potential. The coefficients for
the hopping term and the on-site interaction strength read
\begin{eqnarray}
J &=& - \int {\rm d}x w_l(x) \left( -\frac{\hbar^2 \nabla^2}{2m} + V(x) \right) w_{l+1}(x)\,, \\
U &=& u_{gg}\frac{m\omega_r}{4\pi\hbar} \int {\rm d}x w_l(x)^4\,,
\end{eqnarray}
with the Wannier functions chosen to be real. Note that frozen
transverse dynamics, as assumed in Eq.~(\ref{eqn:wannier}), is
here ensured by taking $J,U\langle {n}\rangle\ll\hbar \omega_r$,
where $\langle {n}\rangle$ is the mean site occupation. Within
this decomposition the term describing atom-light scattering takes
the form

\begin{eqnarray} H_{int}' &=&
\sum_{\lambda,\lambda'}\frac{\hbar C_{\lambda}^*
C_{\lambda'}}{\omega_{\lambda'}-\omega_0}
a_{\lambda}^{\dagger}a_{\lambda'}\mathcal T({\bf q})\,.
\label{eqn:Hint}
\end{eqnarray}
Here,

\begin{equation} \mathcal T
({\bf q})= \sum_l {\rm e}^{{\rm i} q_xld_0}  \left [ J_0(\V{q})
n_l +J_1(\V{q}) \left (  b_l^{\dagger}b_{l+1} +
b_{l+1}^{\dagger}b_l  \right ) \right ] \label{eq:T}
\end{equation}
consists of a photon-dependent energy shift, weighted by the
coefficient

\begin{eqnarray}
J_{0}(\V{q}) &=&
e^{-\frac{1}{4}(q_y^2+q_z^2)\xi_r^2}\int {\rm d}x
{\rm e}^{{\rm i} q_x x} w_0(x)^2\,,
\end{eqnarray}
and a hopping term with coefficient

\begin{eqnarray} J_{1}(\V{q}) &=&
e^{-\frac{1}{4}(q_y^2+q_z^2)\xi_r^2}\int {\rm d}x w_0(x)
{\rm e}^{{\rm i} q_x x} w_0(x-d_0)\,,
\end{eqnarray}
which describes light-assisted tunneling due to the mechanical
effects of photon scattering. This latter term has been neglected
in previous theoretical
treatments~\cite{ReyBraggScattering,Mekhov_Maschler,Meystre}. Its
effect has been investigated in Ref.~\cite{Priscilla,Saba} for
light scattering by ultracold atoms in a double well potential,
showing that the mechanical effect of light can interfere with
ordinary tunneling between the wells, generating observable
effects in the first--order coherence properties of the scattered
light. We hence expect that it will give rise to observable
effects in the Bragg signal by ultracold atoms in optical
lattices.

In the following, we introduce the many-body states, eigenstates
of $H'_g$, which are relevant for the scattering process when the
system is in the Mott-insulator and in the superfluid regime. In
this treatment we use the same notations as in Ref.~\cite{ReyPhd},
and refer the reader to this work for more details, like, e.g.,
the careful comparison between the Bogoliubov approximation and
the exact solution for small one-dimensional systems.

\subsubsection{Mott-insulator State}\label{Sec:MI}

For vanishing hopping,  the ground state of Hamiltonian
(\ref{eqn:Hatomic}) is the Mott-insulator state with all lattice
sites equally occupied with (integer) filling factor $g=N/M$,

\begin{eqnarray}
\ket{\psi_0^{(0)}}
&=&\prod_{l=1}^M\frac{\left(b_l^{\dagger}\right)^g}{\sqrt{g!}}|0\rangle_{\rm
at}= \ket{g,g,...g,g}\,,
\end{eqnarray}
where $|0\rangle_{\rm at}$ denotes the vacuum. The corresponding
ground state energy for $J=0$ is easily found and reads
$E_0^0=MUg(g-1)/2-Mg\mu$. The lowest-lying excitations take the
form
\begin{eqnarray}
\ket{\psi_{n,m}^{(0)}}
&=&\frac{b_n^{\dagger}b_m}{\sqrt{g(g+1)}}|\psi_0^{(0)}\rangle\,,
\end{eqnarray}
where one particle and one hole are created at site
$n$ and $m$, respectively, with energy $E_1^0=E_0^0+U$. These
states form a degenerate subspace of dimension $M(M-1)$. This
degeneracy is lifted for finite values of the hopping $J$.

The corrections due to a non-vanishing but small value of tunneling
are evaluated using perturbation theory. Including the first-order
correction, the ground state now reads

\begin{equation}\label{eqn:Mottgnd}
\ket{\psi_0^{(1)}}=
\left(1-\frac{J^2}{U^2} Mg(g+1)\right)\ket{\psi_0^{(0)}}+\frac{J}{U}\sqrt{2Mg(g+1)}\ket{S}\,,
\end{equation}
where $\ket{S}=\frac{1}{\sqrt{2M}} \sum_n
\left (\ket{\psi_{n,n+1}^{(0)}}+\ket{\psi_{n,n-1}^{(0)}} \right )$ is the
normalized state of adjacent particle-hole excitations, while the
term at second order in $J$ warrants normalization of
state~(\ref{eqn:Mottgnd}). The corresponding energy is
$E_0=E_0^0+{\rm O}(J^2)$. The lowest-lying excitations are
determined using degenerate perturbation theory within the
subspace of single particle-hole excitations,

\begin{equation} \ket{\psi_{[i]}^{(0)}}
=\sum_{n,m}c_{n,m}^{[i]}\ket{\psi_{n,m}^{(0)}}\,,
\end{equation}
where the coefficients $c_{n,m}^{[i]}$ fulfill the normalization
condition and satisfy the equations

\begin{equation}\label{c:nm}
(g+1)(c_{n+1,m}^{[i]}+c_{n-1,m}^{[i]})+g(c_{n,m+1}^{[i]}+c_{n,m-1}^{[i]})=A_i
c_{n,m}^{[i]}\,,
\end{equation}
with periodic boundary conditions

\begin{eqnarray}
c_{n+M,m}^{[i]} &=& c_{n,m+M}^{[i]}=c_{n,m}^{[i]}\,, \label{eqn:c_periodic} \\
c_{n,n}^{[i]} &=&0\,.
\end{eqnarray}
Term $A_i$ in Eq.~(\ref{c:nm}) is the first-order correction to
the corresponding energy, $E_i=E_0+U-J A_i+{\rm O}(J^2)$.

An analytic solution of Eqs.~(\ref{c:nm}) can be derived in the limit of large filling $g\gg1$~\cite{ReyBraggScattering,ReyPhd}. This limit introduces a symmetry between particle and hole excitations, that simplifies the analytical treatment but imposes a selection rule, which is strictly correct only when $g\to \infty$. The coefficients, evaluated in this limit, read

\begin{equation}
\label{eqn:coeff}
c_{n,m}^{[r,s]} =
\frac{\sqrt{2}}{M} \left \{  \begin{array}{l}
                         \sin \left [ \alpha r |n-m| \right ] \,e^{{\rm i} \alpha s (n+m)}  \hspace{0.5cm} \mbox{ for } r+s \mbox{ odd, }\\
                         \sin \left [ \alpha r (n-m) \right ] \,e^{{\rm i} \alpha s (n+m)}  \hspace{0.45cm}\mbox{ for } r+s \mbox{ even, }
                     \end{array}  \right .
\end{equation}
with $\alpha=\frac{\pi}{M}$, $s=0,1...M-1$ and $r=1,2...M-1$. Correspondingly, the lowest-lying excitations and their energy read (at first order in J and for $g\gg 1$)

\begin{widetext}
\begin{subequations}\label{eqn:Ers}
\begin{eqnarray}
\ket{\psi_{[r,s]}^{(1)}} &=&  \left \{ \begin{array}{l}
 \frac{1}{\mathcal N_r} \left ( \sum_{n,m} \left ( c_{n,m}^{[r,0]} \ket{\psi_{n,m}} \right )-\frac{J}{U} \sqrt{8 g (g+1)} \sin \alpha r \ket{\psi_0^{(0)}} \right ) \hspace{0.5cm} \mbox{ for } s=0 \mbox{ and } r \mbox{ odd, } \vspace{0.2cm} \\
                         \sum_{n,m} c_{n,m}^{[r,s]} \ket{\psi_{n,m}} \hspace{0.5cm} \mbox{ otherwise, }
                     \end{array}  \right .  \\ \vspace{0.2cm}
E_{r,s} &=& E_0^0+U-2J (2g+1) \cos \alpha r \cos \alpha s+{\rm
O}(J^2)\,,
\end{eqnarray}
\end{subequations}
\end{widetext}
where $\mathcal N_r$ is a normalization factor. Note that states
$\ket{\psi_{[r,s]}^{(1)}}$ contain a correction proportional to
the ground state $\ket{\psi_0^{(0)}}$. This correction is found
from non-degenerate perturbation theory and warrants the
orthonormality of the new basis
$\{\ket{\psi_0^{(1)}},\ket{\psi_{[r,s]}^{(1)}}\}$.

\subsubsection{Superfluid state} \label{Sec:SF}

In the weakly-interacting superfluid regime, quantum fluctuations
in the number of atoms per site are described by the decomposition

\begin{equation}
\label{eqn:bog} b_l=z_l+\beta_l\,,
\end{equation}
where $z_l$ is a complex number describing the order parameter and
$\beta_l$ is the fluctuations operator obeying the bosonic
commutation rules. The order parameter $z_l$ is found by
minimizing Hamiltonian~(\ref{eqn:Hatomic}) at zero-th order in the
expansion in $ \beta_l$ and $\beta_l^{\dagger}$. It obeys the
discrete nonlinear Schr\"odinger equation

\begin{equation}
\label{eqn:DNLSE} \mu z_l = -\sum_{\langle k,m\rangle} J_k z_m
\delta_{k,l} + U|z_l|^2 z_l\,,
\end{equation}
where $|z_l|^2$ corresponds to the condensate fraction. For a
translationally invariant lattice and in the limit of weak
interactions, as considered here, it is given by $|z_l|^2=g$, and
Eq.~(\ref{eqn:DNLSE}) reduces to

\begin{equation}\label{eqn:chempot}
\mu = -2J + U g\,.
\end{equation}
Using Eq.~(\ref{eqn:bog}) and Eq.~(\ref{eqn:DNLSE}) in
Hamiltonian~(\ref{eqn:Hatomic}), keeping only terms up to second
order in the operators $\beta_l$, $\beta_l^{\dagger}$, one finds
$H_{g}'=H_0+H_2$ with

\begin{eqnarray} H_2 =  \sum_{l,m} \mathcal
L_{l,m} \beta_l^{\dagger}\beta_m +\mathcal M_{l,m}
\beta_l^{\dagger}\beta_m^{\dagger} + {\rm H.c.}\,,
\end{eqnarray}
where the coefficients read
\begin{eqnarray}
\mathcal L_{l,m} &=& -J\sum_{\langle n,k\rangle} \delta_{n,l}\delta_{m,k}/2 +
\delta_{l,m}(2Ug -\mu)/2\, \nonumber \\
\mathcal M_{l,m} &=& U g \delta_{l,m}/2\,. \nonumber
\end{eqnarray}
Term $H_2$ describes the dynamics of the non-condensed fraction at
leading order. It can be written in the diagonal form

\begin{equation}
H_2 = \sum_{p\ne 0} \hbar \Omega_p \left (
\alpha_p^{\dagger} \alpha_p+\frac{1}{2} \right )-\sum_l \mathcal
L_{ll}\,,
\end{equation}
where operators  $\alpha_p$ and $\alpha_p^{\dagger}$ are
respectively the bosonic annihilation and creation operators of
the Bogoliubov excitation with quasimomentum  $p=n2\pi/Md_0$, with
$n=-M,M-1,\ldots, M-1$, and the frequency $\Omega_p$ is given by

\begin{eqnarray}
\hbar^2\Omega_p^2 &=& \epsilon_p^2+2Ug\epsilon_p\,,
\label{eqn:bog_disp}
\end{eqnarray}
with
\begin{equation}
\epsilon_p = 4J \sin^2 \left (\frac{d_0p}{2} \right )\,.
\end{equation}
Operators $\alpha_p$, $\alpha_p^{\dagger}$ satisfy the
commutation relations $\left[ \alpha_p,\alpha_{p'}^{\dagger}
\right ]=\delta_{p,p'}$ and are related to $\beta_l$ by the
Bogoliubov transformation

\begin{eqnarray}
\beta_l &=& \frac{1}{\sqrt{M}} \sum_{p\ne 0} e^{{\rm i} pld_0}u_p
\alpha_p-e^{-{\rm i} pld_0}v_p\alpha_p^{\dagger}. \label{eqn:bogA}
\end{eqnarray}
The Bogoliubov amplitudes $u_p$, $v_p$ satisfy the equation
$|u_p|^2-|v_p|^2=1$, as a consequence of the commutation
relations, and depend only on the modulus of the quasimomentum,
$u_p=u_{-p}$, $v_p=v_{-p}$. They are solutions of the Bogoliubov-de Gennes equations, which in our case read

\begin{equation} \left ( \begin{array}{cc}
                        J \epsilon_p + U g & -U g  \\
                        U g & -J \epsilon_p -U g
                     \end{array}  \right ) \left ( \begin{array}{c}
                        u_p  \\
                        v_p
                     \end{array}  \right ) =\hbar \Omega_p \left ( \begin{array}{c}
                        u_p  \\
                        v_p
                     \end{array}  \right ).
\end{equation}
In particular,

\begin{subequations}\label{eqn:BogDeGennesLsg} \begin{eqnarray}
u_p^2 &=& \frac{\epsilon_p+Ug+\hbar \Omega_p}{2\hbar \Omega_p}\,, \\
v_p^2 &=&  \frac{\epsilon_p+Ug-\hbar \Omega_p}{2 \hbar \Omega_p}\,, \\
u_p v_p &=& \frac{U g}{2 \hbar \Omega_p}\,. \end{eqnarray}
\end{subequations}

We note that $\epsilon_p$ is the energy of a non interacting
particle in the lattice. By replacing it with the free-space
energy $\epsilon_p\rightarrow p^2/2m$ we recover in
Eq.~(\ref{eqn:bog_disp}) the dispersion relation for a
weakly-interacting dilute Bose gas in free space \cite{Stringari}.
Contrarily to the case of the uniform one-dimensional system,
where Bogoliubov theory is not applicable, for a finite system
it provides a well defined and small depletion for $U/J\ll 1$ and large filling. The
corresponding spectrum of the differential scattering cross
section will be compared below with the numerical results obtained
for a finite Bose-Hubbard model composed of 7 atoms.

In the following, we will denote by $\ket{0}_{\rm SF}$ the
superfluid state, where all atoms are in the condensate, and by
$\ket{p}_{\rm SF}=\alpha_p^{\dagger}|0\rangle_{\rm SF}$ the state
with one Bogoliubov excitation at quasimomentum $p$. In
particular, we will consider scattering processes, such that the
state of the atoms will include at most one Bogoliubov excitation.
To this aim, it is convenient to rewrite operator $\mathcal T({\bf
q})$ in Eq.~(\ref{eq:T}) using the decomposition of operator $
b_l$ in Eq.~(\ref{eqn:bog}),

\begin{equation} \mathcal T_{\rm SF�}({\bf
q})=\mathcal T_{\rm SF}^{(0)}({\bf q})+\mathcal T_{\rm
SF}^{(1)}({\bf q})+\mathcal T_{\rm SF}^{(2)}({\bf q})\,,
\end{equation}
The first term on the right-hand side of the equation
describes radiation coupling with the condensate and reads

\begin{equation} \mathcal T_{\rm SF}^{(0)}({\bf
q})=g\left(J_0({\bf q})+2J_1({\bf q}) \right)\sum_l{\rm e}^{{\rm i}q_xld_0}\,,
\end{equation}
while the other terms give radiation coupling with the Bogoliubov
excitations, and take the form

\begin{eqnarray}
\mathcal T_{\rm
SF}^{(1)}({\bf q})&=&\sqrt{g}\sum_l{\rm e}^{{\rm
i}q_xld_0}\left(J_0({\bf q})(\beta_l+\beta_l^{\dagger})\right.\\
& &\left.+J_1({\bf q})(\beta_l^{\dagger}+\beta_{l+1}+\beta_{l+1}^{\dagger}+\beta_{l})\right)\,, \nonumber\\
\mathcal T_{\rm SF}^{(2)}({\bf q})&=&\sum_l{\rm e}^{{\rm
i}q_xld_0}\left(J_0({\bf
q})\beta_l^{\dagger}\beta_l\right.\\
& &\left.+J_1({\bf q})(\beta_l^{\dagger}\beta_{l+1}+
\beta_{l+1}^{\dagger}\beta_{l}) \right)\,,\nonumber
\end{eqnarray}
where the superscript gives the order in the Bogoliubov expansion.

\section{Light scattering} \label{Sec:3}

Light scattering by a one-dimensional optical lattice of ultracold
atoms is studied in  the setup sketched in Fig.~\ref{fig:setup}. A
laser plane wave at wave vector ${\bf k_L}$, frequency
$\omega_L=c|{\bf k_L}|$, in a coherent state with amplitude
$\alpha_L$, drives the atoms. We evaluate the scattered light as a
function of the angle of emission, determined by the wave vector
${\bf k}$ of the mode into which the photon is emitted, and of the
frequency of the emitted photon.

The scattering process is evaluated assuming that the laser very
weakly excites the atom, so that the atom-photon interaction is
described at lowest order by Hamiltonian~(\ref{eqn:Heff}). More in
detail, the condition $|\alpha_L|\ll 1$ means that the atomic
sample is driven by at most one photon. A scattering process will
then occur with probability $|\alpha_L|^2$ and  will consist of
the absorption of one incident photon in the mode of the laser,
represented by the state $|1_L\rangle$, and the emission of a
photon in one of the modes of the electromagnetic field at wave
vector ${\bf k}$ and polarization $\epsilon_{\bf k}\perp {\bf k}$,
represented by the state $|1_{{\bf k},\epsilon}\rangle$. The
corresponding differential scattering cross section for the photon
scattered at frequency $\omega$ in direction ${\bf n}$ in the
solid angle $\Omega$ is proportional to the scattering
rate~(\ref{rate}) and takes the form~\cite{Cohen,Footnote}

\begin{widetext}
\begin{equation} \label{cross:section}
\sigma(\Omega,\omega)= \frac{\mathcal V^2\omega_L^2}{(2\pi)^2
\hbar ^2 c^4} \sum_f \sum_{\epsilon_{\bf k}\perp {\bf n}}\left
|\bra{f,1_{{\bf k},\epsilon}}H_{\rm int}' \ket{i,1_L} \right |^2
\delta^{(T)}(\omega_L+\omega_i-\omega-\omega_f)\,,
\end{equation}
\end{widetext}
where ${\bf k}={\bf n} k$ and $\ket{i},\ket{f}$ are the initial
and final atomic states, eigenstates of
Hamiltonian~(\ref{eqn:Hatomic}) at the eigenfrequencies $\omega_i$
and $\omega_f$, respectively. Using Eq.~(\ref{eqn:Heff}) in
Eq.~(\ref{cross:section}) one can easily verify that the
differential scattering cross section is proportional to the
dynamic structure factor~\cite{ReyBraggScattering}.

We evaluate the scattering cross section assuming that the atoms
are initially in the ground state either of the Mott-insulator or
of the superfluid phase, and that the atoms are scattered into a
final state belonging to the lowest-lying atomic excitations.
Using the form of operator $H_{\rm int}'$ in Eq.~(\ref{eqn:Hint}),
Eq.~(\ref{cross:section}) can be written as

\begin{equation}
\sigma(\Omega,\omega)=\sigma^{(0)}(\Omega,\omega)+\sigma^{(1)}(\Omega,\omega)\,,
\end{equation}
where

\begin{equation}\sigma^{(0)}=\mathcal
A(\Omega) \left |\bra{i}\mathcal T({\bf q}) \ket{i} \right |^2
\delta^{(T)}(\omega_L-\omega)
\end{equation}
gives the elastic component of the scattered light, while
\begin{eqnarray}
\sigma^{(1)} &=&\mathcal A(\Omega)\sum_f\left |\langle f,i_{\bf
k}|\mathcal T({\bf q}) \ket{i,1_L} \right
|^2\delta^{(T)}(\omega_L-\omega-\delta\omega_f) \nonumber \\
\end{eqnarray} describes the scattering events in which one
mechanical excitation at frequency $\delta\omega_f$ is absorbed
from the photon by the atomic lattice (Stokes component) and
corresponds to the one-phonon terms in neutron
scattering~\cite{Ashcroft}. The corresponding phonon emission
processes, giving the anti-Stokes component, are here absent as
initially the atoms are in the ground state. Moreover, higher
order terms, corresponding to higher-order phonon terms in neutron
scattering, are here neglected as we assume that at most one
mechanical excitation is exchanged between lattice and photons.

The operator $\mathcal T({\bf q})$ in the above equations is given
in Eq.~(\ref{eq:T}), while the coefficient $\mathcal A(\Omega)$
depends on the angle of emission and takes the form

\begin{eqnarray}
\mathcal A(\Omega) &=&\frac{\mathcal
V^2\omega_L^2}{(2\pi)^2\epsilon_0^2\hbar^2c^4}\sum_{\epsilon_{\bf
k}\perp {\bf n}}\frac{\hbar^2|C_LC_{\bf
k}|^2}{|\omega_L-\omega_0|^2}\nonumber\\
&=&\frac{\gamma}{c}
\frac{\Omega_0^2}{\Delta^2}\left[\frac{3}{8\pi}\left(1-\frac{|{\bf
D}\cdot {\bf n}|^2}{|{\bf D}|^2}\right)\right]\,,
 \end{eqnarray}
where $\gamma$ is the linewidth of the dipole transition,
$\Delta=\omega_L-\omega_0$ is the detuning of the laser from the
atomic transition and
$\Omega_0=\sqrt{\omega_L/2\hbar\varepsilon_0}{\bf D}\cdot
\epsilon_L$.

\subsection{Scattering cross section as a function of the atomic
state}

We now give an analytic expression for the scattering cross
section in Eq.~(\ref{cross:section}) for the initial and final
states determined in Sec.~\ref{Sec:MI} and~\ref{Sec:SF}.

\subsubsection{Mott-insulator}

For the Mott-insulator phase the initial state is
$|i\rangle=|\psi_0^{(1)}\rangle$ given in Eq.~(\ref{eqn:Mottgnd}).
Using Eq.~(\ref{eqn:Hint}), we find

\begin{eqnarray}
&&\sigma^{(0)}_{\rm MI}(\Omega,\omega) =\mathcal A(\Omega)
N^2\delta(\omega_L-\omega) \delta_{q_x,G}^{(M)}
\label{W:MI}\\
&&\times  \left(|J_0(\V{q})|^2+4\sqrt{g(g+1)}\frac{J}{U}{\rm
Re}\left\{J_0^*({\bf q})J_1({\bf q}) \right\}\right),  \nonumber
\end{eqnarray}
where $G$ are the vectors of the (one-dimensional) reciprocal
lattice and
\begin{eqnarray}
\delta_{q,G}^{(M)}\equiv \frac{1}{M^2} \frac{\sin^2(Md_0q/2)}{\sin^2(d_0 q/2)}
\end{eqnarray}
gives conservation of the Bloch momentum in a finite lattice with
$M$ sites, such that $\delta_{q,G}^{(M)}\to \delta_{q,G}$
(Kronecker delta) as $M\to \infty$. In Eq.~(\ref{W:MI}) we omitted
terms at third and higher order in $J$ and $J_1({\bf q})$. This
approximation will be applied to the rest of this section,
assuming that these higher-order terms can be neglected.

The presence of $\delta_{q_x,G}^{(M)}$ in Eq.~(\ref{W:MI})
expresses the von-Laue condition for Bragg scattering. At zero
order in the hopping term, Eq.~(\ref{W:MI}) gives the response of
a crystal of particles oscillating around their equilibrium
position. In fact, using a Gaussian ansatz for the wave functions,
one can estimate $|J_0(\V{q})|^2 \simeq {\rm e}^{-2W}$, with
$W=[q_x^2 \xi_x^2+(q_y^2+q_z^2)\xi_r^2]/8$, where $\xi_x$ and
$\xi_r$ are the widths the atomic wave functions in the axial and
radial direction, showing explicitly that this term is analogous
to the Debye-Waller factor~\cite{Ashcroft,Kohn}. The term proportional
to $J$ is instead a novel feature with respect to
traditional condensed-matter systems, that arises from light
induced tunneling.

The Stokes component for the Mott-insulator is evaluated taking
the final states $|f\rangle=|\psi_{[r,s]}^{(1)}\rangle$ given in
Eq.~(\ref{eqn:Ers}), and reads

\begin{eqnarray}
\label{sigma:1}\sigma_{\rm MI}^{(1)}(\Omega,\omega)&=&\mathcal
A(\Omega)  \sum_{r,s} \sin^2\left(\frac{\pi r}{M}\right)|\mathcal B_{r,s}|^2 \\
&\times &\delta(\omega_L-\omega-\omega_{r,s})\delta_{q(s),G}^{(M)}\,,
\nonumber
\end{eqnarray}
with $\omega_{r,s}=(E_{r,s}-E_0)/\hbar$, and where we have
introduced

\begin{equation} \label{eqn:qs} q(s)=q_x-\frac{2\pi}{Md_0}s\,. \end{equation}
The coefficient in Eq.~(\ref{sigma:1}) takes the form

\begin{equation}\label{eqn:Ars}
\mathcal B_{r,s} = \sqrt{8g(g+1)}\left \{  \begin{array}{l}
                        J_1(\V{q})  \hspace{0.2cm} \mbox{ for } r+s \mbox{ odd, } \vspace{0.4cm} \\
                        2 \frac{J}{U}J_0(\V{q})\sin \left(\frac{\pi}{M}s\right)     \hspace{0.2cm} \mbox{ for } r+s \mbox{ even, }
                     \end{array}  \right .
\end{equation}
showing that the transition to the excited states with $r+s$ odd
is  due to photon recoil, and is hence a light-induced hopping
process. Note that condition $q(s)=G$ shows that the quantum
number $s$, and more specifically $2\pi s/L$, with $L=Md_0$ the
length of the lattice, plays the role of the quasi-momentum of the
states $\ket{\psi_{r,s}^{(1)}}$. We remark that
Eq.~(\ref{eqn:Ars}), for $r+s$ even, agrees with the result
evaluated in~\cite{ReyBraggScattering} (see Eq.~(9) of that paper
for comparison). The result we find for $r+s$ odd, on the
contrary, is discarded in the treatment
of~\cite{ReyBraggScattering}, as there the authors neglected light
induced hopping terms. In the Mott-insulator regime these terms
are usually very small with respect to the other contributions.
They give rise to a significant contribution when interfering with
ordinary tunneling. This latter type of contributions is ruled out
in the analytical model by the selection rule introduced by the
assumption $g\gg 1$, but it is visible in the numerical results,
as it will be shown in Sec.~\ref{Sec:3}.

\subsubsection{Superfluid}

When evaluating the differential scattering cross section in the
superfluid phase, we assume all atoms to be initially prepared in
the Bose-Einstein condensate. In addition, for the analytical calculation we consider the limit $U\rightarrow 0$. In this limit we can neglect the quantum depletion of the condensate due to the interactions and take the initial state
$|i\rangle=|0\rangle_{\rm SF}$ according to our notation. The zero-phonon term takes now the form

\begin{widetext}
\begin{eqnarray}\label{W:SF} &&\sigma_{\rm SF}^{(0)}(\Omega,\omega) =\mathcal
A(\Omega) \delta(\omega_L-\omega) \delta_{q_x,G}^{(M)}N^2\left( \left|J_0(\V{q})+2J_1({\bf q})\right|^2 +2\sum_{p\neq 0}\frac{|v_p|^2}{N}{\rm Re}\left\{(J_0(\V{q})+2J_1({\bf q}))^*(J_0(\V{q})+2J_1({\bf q})\cos \left( pd_0)\right )\right\}\right)\,, \nonumber \\
\end{eqnarray}
showing that the light-induced tunneling effects enter already at first order
in this expression. As in the Mott-insulator case, the analogous
of the Debye-Waller factor can be here identified in the term
$|J_0({\bf q})|^2$. In this case, though, tunneling effects
become more important, modifying significantly the signal
as we will show. The first-phonon term reads

\begin{eqnarray}\sigma_{\rm SF}^{(1)}(\Omega,\omega) =\mathcal
A(\Omega) N \sum_{p\ne 0} \delta(\omega_L-\omega-\Omega_p) \frac{\epsilon_p}{\hbar \Omega_p} \left| \left(J_0({\bf q})+J_1({\bf q})(1+{\rm e}^{-{\rm i}pd_0})\right) \right|^2  \delta_{q_x-p,G}^{(M)}\,, \end{eqnarray}
\end{widetext}
and  describes the creation of Bogoliubov excitations with
quasi-momentum $\hbar p$ by photon scattering, such that the
relation $p=q_x-G$ holds.

\subsection{Numerical results} \label{Sec:Numerics}

In this section we report the numerical results for the
differential scattering cross section obtained when the atoms are
in the Mott-insulator or in the superfluid state.  The numerical results
are obtained for a lattice of $M=7$ sites and fixed particle
number $N=M$. The coefficient entering the Bose-Hubbard
Hamiltonian in Eq.~(\ref{eqn:Hatomic}) and the operator $\mathcal
T ({\bf q})$ in Eq.~(\ref{eq:T}) are calculated by using the Wannier
functions relative to a given lattice depth $V_0$ of optical
potential~(\ref{pot}). Hamiltonian~(\ref{eqn:Hatomic}) is
diagonalized exactly and the corresponding states are used for
determining the differential scattering cross section in
Eq.~(\ref{cross:section}). The numerical results are also compared
with the analytical predictions of the scattering cross sections
reported in the previous section. Although the latter are valid
for very large lattices and for large mean site occupation $g\gg
1$, we find reasonable agreement when comparing these predictions with
those for a small lattice of 7 sites and single occupancy (see
also~\cite{ReyBraggScattering}). 
\begin{figure*}[htp] \centering
\includegraphics[width=\textwidth]{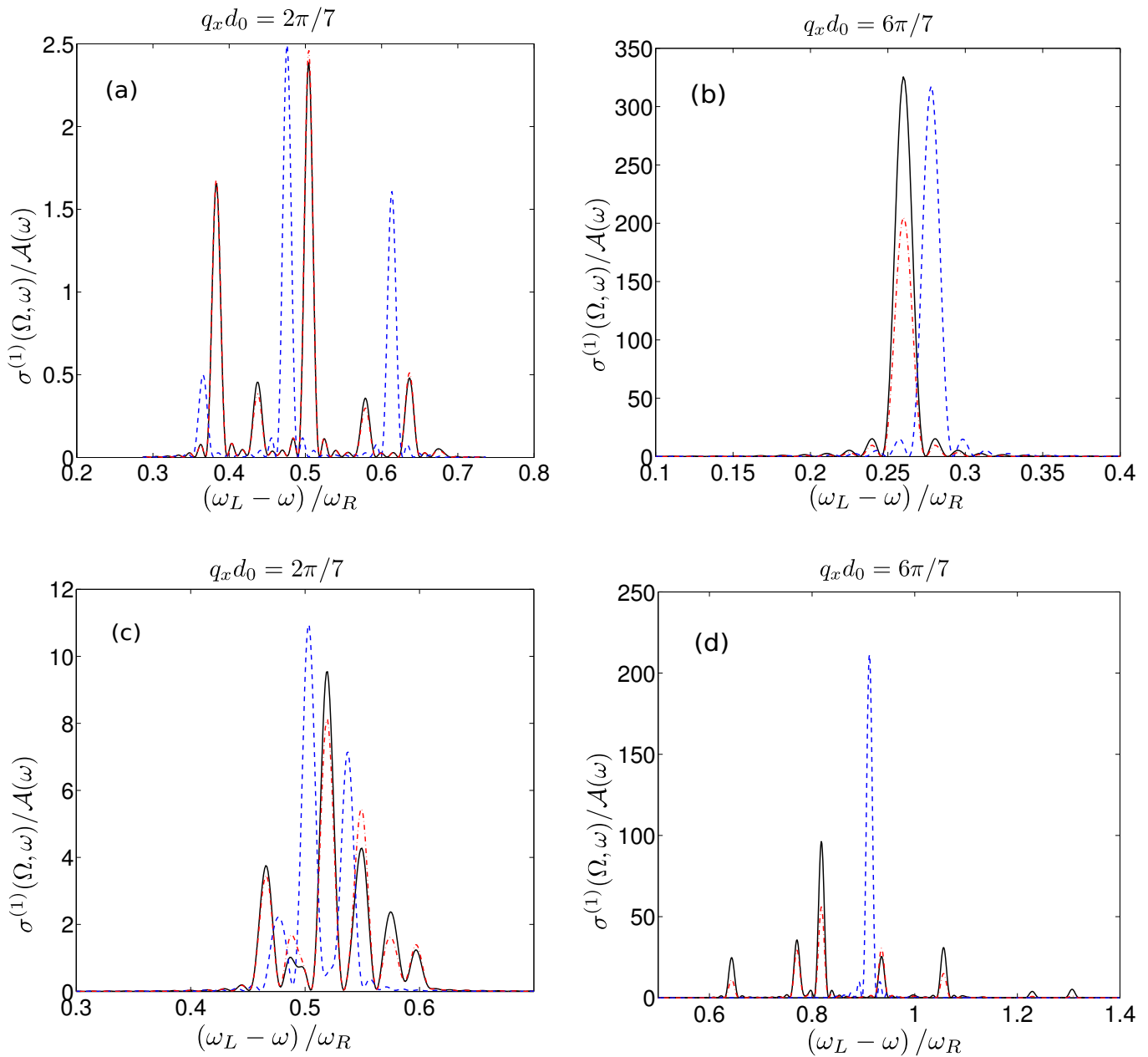}
 \caption{\label{fig:spectrum} (color online) Stokes
component of the differential scattering cross section  (in units
of $\mathcal A(\Omega)$) as a function of frequency (in units of
the recoil frequency $\omega_R$) for two different scattering
angles, corresponding to $q_xd_0=2\pi/7$ (top row) and to
$q_xd_0=6\pi/7$ (bottom row). The curves have been evaluated for a
lattice of $M=7$ site and $N=M=7$ composed by $^{87}$Rb atoms in
the $\ket{{\rm F}=2,m_F=2}$ hyperfine ground state. The black
solid line corresponds to the numerical results, the blue dashed
line to the analytical formulas (see text), the red dashed-dotted
line to the model of~\cite{ReyBraggScattering}, where the
light-induced hopping is neglected. Plots (a) and~(c) are
evaluated for $V_0 =8.1 \hbar\omega_R$ ($U/J \approx 17$) which
corresponds to the Mott-insulator state. Plots (b) and~(d) are
evaluated for $V_0 =0.1 \hbar\omega_R$ ( $U/J \approx 1$) which
corresponds to the superfluid state. Other parameters are
$d_0=413$nm, $a_s=105a_0$ with $a_0$ being the Bohr radius, and $ 
\omega_r = 10 \omega_R$ corresponding to the experimental 
parameters in~\cite{Stoeferle} (For these parameters the size of
the radial wavepacket is $\xi_r=10a_s$). The frequency resolution
is set to $\Delta \omega = 300$~Hz, corresponding to an
integration time $T=3$~msec. } \end{figure*}

Figure~\ref{fig:spectrum}(a) and~(c) display the one-phonon
contribution to the differential scattering cross section,
$\sigma^{(1)}(\Omega,\omega)$, as a function of the frequency
$\omega$ and for different scattering angles when the atoms are in
the Mott-insulator state. The numerical results are compared with
the analytical model (dashed line) and with the model used in the
numerical simulations in~\cite{ReyBraggScattering}, in which
light-induced hopping terms are not considered.
\begin{figure*}[htp]
\centering
\includegraphics[width=\textwidth]{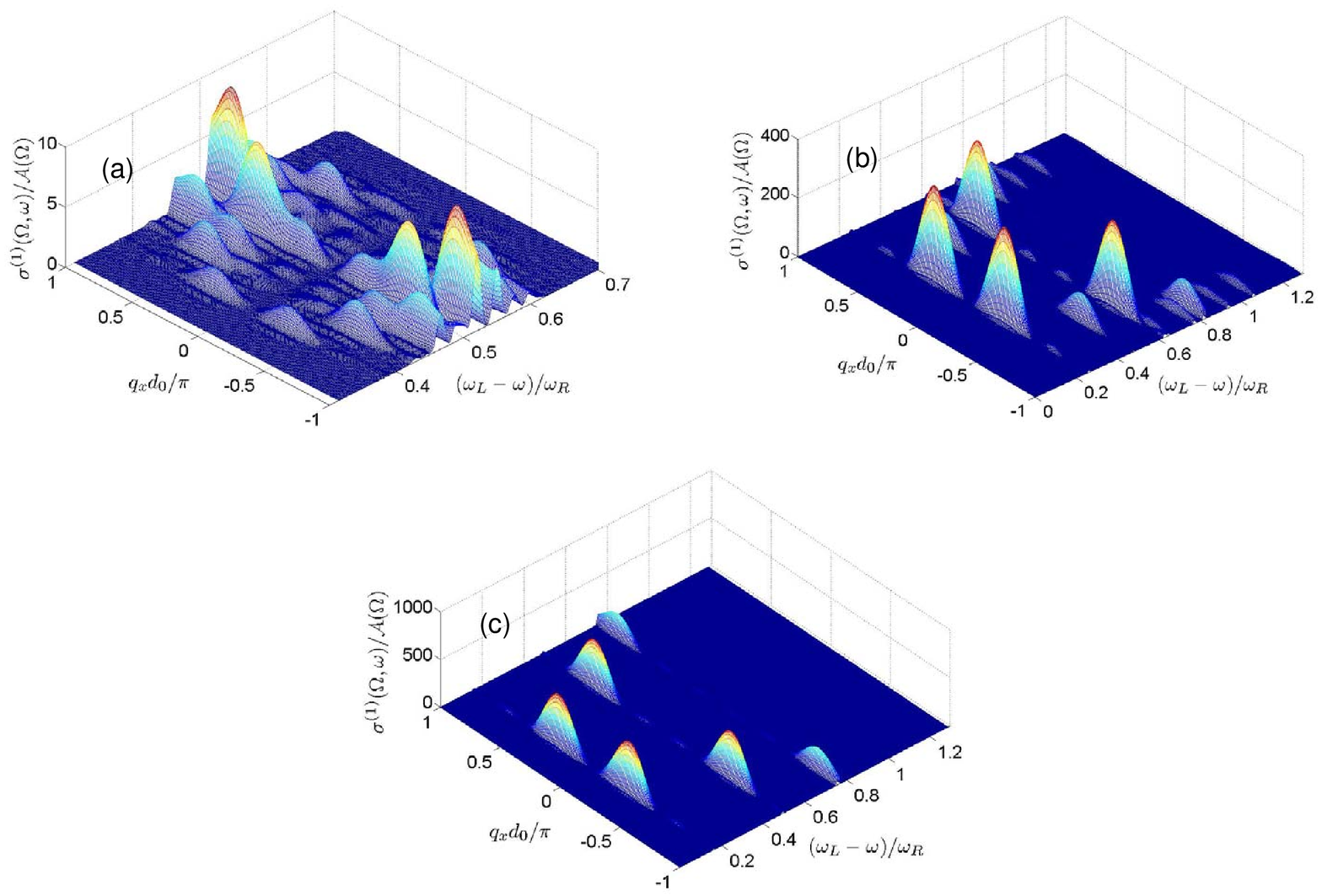}
\caption{ \label{fig:spectrum3D} (color online) Stokes component of
the differential scattering cross section (in units of $\mathcal
A(\Omega)$) as a function of the frequency (in units of
$\omega_R$) and of the Bragg angle $\Theta=q_xd_0$ (in units of
$\pi$). The plots have been evaluated numerically for (a) $V_0
=8.1 \hbar \omega_R$ and $U/J \approx 17$, (b) $V_0 =0.1
\hbar\omega_R $ and $U/J \approx 1$, (c) $V_0=0.1 \hbar\omega_R$
and $U/J \approx 0.1$. The other parameters are as in
Fig.~\ref{fig:spectrum}.}\end{figure*}

We first discuss the numerical results which most closely approach
the exact solution. The appearance of multiple peaks corresponds
to the excitations of the atoms in the Mott-insulator due to the
photon recoil. The number of peaks for the numerical result is
$M-1$, which correspond in this case to 6. They can be
individually resolved, as the system considered here is finite,
and the width of each individual peak is limited by the detection
time $T$ (or the spectral resolution
$1/T$)~\cite{Footnote:Linewidth}. The analytical results are found
using the model described in Sec.~\ref{Sec:2}, which assumes a
large on-site occupation. They are characterized by the same peak
number, although only half of them is visible in the figure. In
fact, the intensity of the peaks arising from the coupling of the
ground state to the corresponding excitation via light-induced
hopping (corresponding to the terms in Eq.~(\ref{eqn:Ars}) with
$r+s$ odd) are very small compared to the other ones
(corresponding to the terms with $r+s$ even) and are therefore not
visible (note that, due to the assumption of large on-site
occupation, interference between ordinary tunneling and
light-induced hopping is suppressed). The central positions of the
visible peaks present a systematic shift with respect to the ones
found numerically. This systematic shift originates from the
assumption $g\gg1$, and has been observed
in~\cite{ReyBraggScattering}. Nevertheless, the analytical
solution still provides some insight into the numerical results.
In Eq.~(\ref{eqn:Ers}), using Eq.~(\ref{eqn:qs}) we find that the
peaks are centered around the energy $E'=U$ with a spreading about
this mean value of width $4J(2g+1)\cos \left (\frac{q_x d_0}{2}\right )$. Such
spreading decreases as $q_x d_0$ approaches $\pi$, compare
Fig.~\ref{fig:spectrum}(a) and~(c). In particular, for
$q_xd_0=\pi$, the width of the distribution of the Stokes
excitations vanishes and the spectrum reduces to a single peak,
corresponding to the on-site energy $U$.

The results for the superfluid regime are reported in
Figs.~\ref{fig:spectrum}(b) and~(d). Here, the analytical solution
predicts that in the limit $g\gg 1$ the total momentum of photon
and lattice is conserved in a scattering event. Such property
implies that the Bogoliubov mode matching the
momentum-conservation condition, is excited, and therefore one
expects a single peak in the spectrum. For $N=7$ atoms and $g=1$,
the numerical results for $U/J\approx 1$ give a single peak at
$q_x d_0=2\pi/7$, while at $q_x d_0=6\pi/7$ multiple peaks are
found. In this case, instead of a collective density fluctuation
with a well defined momentum $p$, one observes particle-hole types
of excitations as in the Mott-insulator case. In
Fig.~\ref{fig:spectrum}(d) one observes a larger spread of the
peaks as compared to the Mott-insulator case at the same Bragg
angle. This is due to the larger value of the tunneling rate $J$.
We remark that, choosing smaller values of the ratio $U/J$ by
ramping down the on-site interaction strength, as it is shown
below, the spectrum reduces to a single peak at all Bragg angles
and approaches the limit of the single-particle spectrum, as it is
recovered in Eq.~(\ref{eqn:bog_disp}) by setting $U=0$.

\begin{figure}[htp]\centering
\includegraphics[width=.44\textwidth]{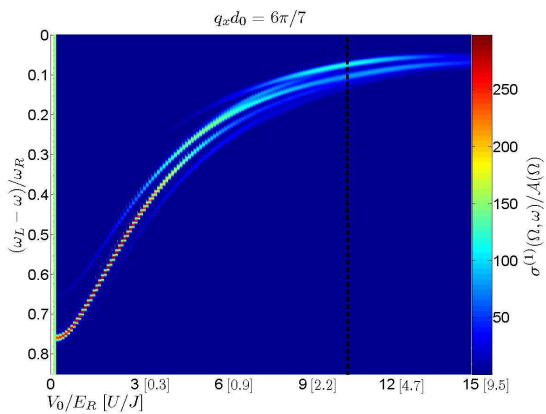}
\caption{\label{fig:SFtoMott} (color online) Contour plot of the Stokes component
of the differential scattering cross section (in units of
$\mathcal A(\Omega)$) as a function of the frequency (in units of
$\omega_R$) and of the lattice depth $V_0$  in units of $\omega_R$
for $q_xd_0=6\pi/7$ (the corresponding value of the ratio $U/J$ is
reported in the axis between squared bracket). The black dashed line marks the critical value at which the phase transition occurs in the thermodynamic limit. The other parameters are given in Fig.~\ref{fig:spectrum}. }\end{figure}

We now compare the numerical results, obtained taking 
systematically into account the light-induced hopping term, to the
results found when this term is neglected, corresponding to the
treatment in~\cite{ReyBraggScattering}. In the Mott-insulator
case, comparison between the numerical results with and without
light-induced hopping effects shows that in the first case one
finds interference between ordinary tunneling and light-induced
hopping. This gives rise to an alternating enhancement and
reduction of the peak heights at different frequencies, which is 
absent in the model discarding light-induced hopping effects. In
general, the light-hopping term contributes in determining the  
height of some peaks, giving substantial modifications of the
spectrum which can be revealed experimentally. The effect is
larger in the superfluid regime, where tunneling is enhanced, as
one can see in Fig.~\ref{fig:spectrum}(b). Here, the central peak
at $q_xd_0=2\pi/7$ is 50\% higher than in absence of this 
contribution.

\begin{figure*}[htp]
\centering
\includegraphics[width=\textwidth]{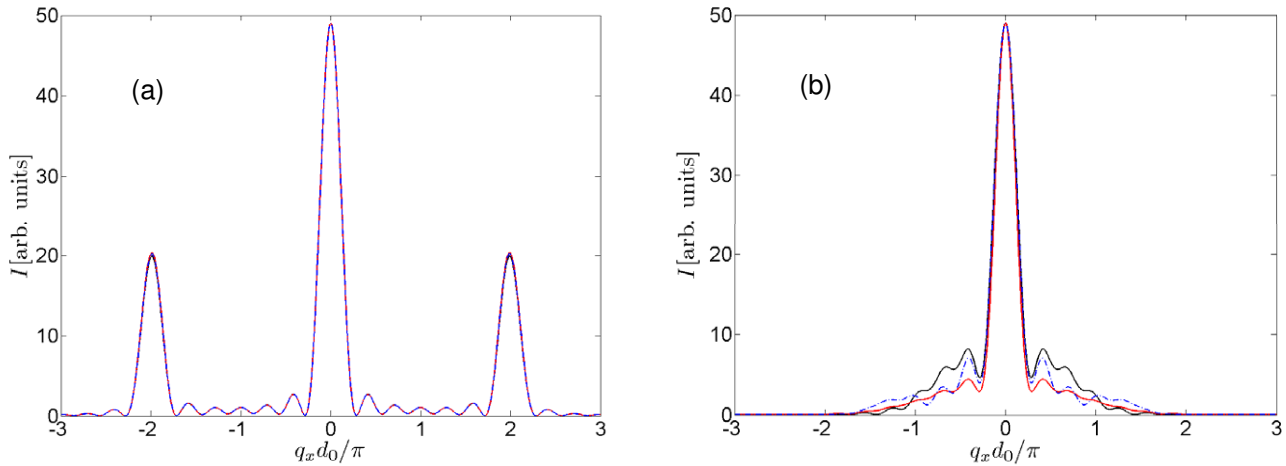}
\caption{\label{fig:Intensity} (color online) Intensity of the
scattered light (in arbitrary units) as a function of the Bragg
angle $\Theta=q_x d_0$ (in units of $\pi$). The parameters are the
same as in Fig.~\ref{fig:spectrum} and (a) $V_0 =8.1 E_R$ ( $U/J
\approx 17$ ), (b) $V_0 =0.1 E_R$ ( $ U/J \approx 1$). The black
solid line corresponds to the numerical result, the blue
dashed-dotted line to the analytical solution, the red dashed line
to the numerical result obtained discarding the light-induced
hopping term as  in~\cite{Mekhov_Maschler,ReyBraggScattering}. } 
\end{figure*}

Figures~\ref{fig:spectrum3D}(a)-(c) display the spectra of
$\sigma^{(1)}$ as a function of the frequency and of the Bragg
angle, in three different points of the phase diagram. We remark
that the width and spacing of the Bragg peaks are determined by
the finite size of the lattice. The plots in~(a) and~(b) are made
in the same parameter regimes as in
Fig.~\ref{fig:spectrum}~(a),(c) and (b),(d), respectively, , 
namely $U/J \approx 17$ and $U/J \approx 1$.
Figure~\ref{fig:spectrum3D}(c), instead, corresponds to the value
$U/J\sim 0.1$. Here, one observes almost a single peak at each
Bragg angle, as expected in the weakly-interacting superfluid
phase.

Figure~\ref{fig:SFtoMott} shows $\sigma^{(1)}$ as a function of
the frequency and the depth of the potential, hence sweeping from
the Mott-insulator to the superfluid regime at a given Bragg
angle, corresponding to large momentum transfer ($q_xd_0=6\pi/7$).
Here, one observes that the spectrum varies from multiple peaks,
deep in the Mott-insulator regime, to a single peak in the
weakly-interacting superfluid regime. The single peak appears
around a value of $U/J$ much smaller than the critical value
$[U/J]_c$ for the Mott to superfluid transition (which, in the
thermodynamic limit, is predicted for $[U/J]_c = 3.37$, see
Ref.~\cite{Monien}). The presence of multiple peaks also in the superfluid phase close to the phase transition is reminiscent of a strongly-interacting superfluid phase. Such phase contains, beyond
the gapless phononic modes, also gapped
modes~\cite{sengupta_pra2005,nikuni,ohashi_pra2006,huber_prb2007,
menotti2008}, which are predicted to be dominant at large
quasi-momentum. We expect that also in the thermodynamic limit the
transition to a single peak in the scattered-light spectrum will
occur at lower values of $U/J$ than the Mott-insulator to
superfluid phase transition and will be also dependent on the
momentum transfer. The identification of the Mott-insulator to
superfluid phase transition should rather rely on the existence of
a gapless spectrum. In spite of the very small size of the
considered system, indications of a gapless spectrum are present
in our results, as one can see comparing
Fig.~\ref{fig:spectrum3D}(a) with (b),(c).

\begin{figure}[htp] \centering
\includegraphics[width=.44\textwidth]{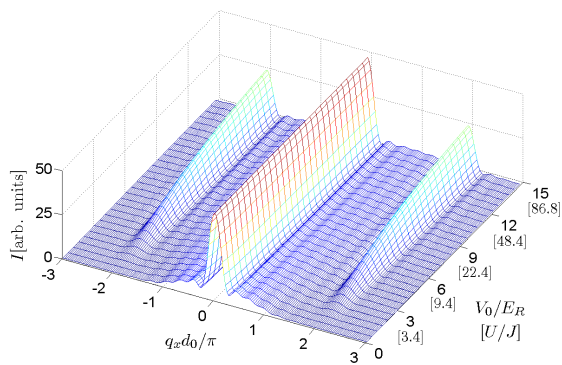}
\caption{\label{fig:Intensity:1} (color online) Intensity of the scattered light
(in arbitrary units) as a function of the Bragg angle $\Theta=q_x
d_0$ (in units of $\pi$) and of the lattice depth $V_0$ (in units
of $\hbar\omega_R$) (the corresponding values of the ratio $U/J$
are reported between squared brackets). The other parameters are
reported in Fig.~\ref{fig:spectrum}. }
\end{figure}

The intensity of the scattered light as a function of the Bragg
angle is determined by the differential scattering cross section

\begin{equation}
\frac{d\sigma}{d \Omega} =\int {\rm d}\omega \sigma(\Omega,\omega)\,,
\end{equation}
and is reported in Figs.~\ref{fig:Intensity} for the atoms in (a) the
Mott-insulator and in (b) the superfluid state. The solid line
here corresponds to the numerical results, the dashed line to the
analytical predictions and the dashed-dotted line to the model
where light-induced hopping  has been discarded, similar to the
case considered in Ref.~\cite{Mekhov_Maschler}. In this latter
work, in fact, corrections due to the tunneling $J$ were neglected
when evaluating light scattering by the atoms in the
Mott-insulator state, while the calculation of light scattering
from the superfluid state was made discarding the finite value of
the on-site interaction as well as the finite width of the Wannier
functions. In Fig.~\ref{fig:Intensity}(a) we observe that in the
Mott-insulator regime the signal is dominated by the elastic
component, and corresponds to a classical diffraction grating. In
the superfluid regime, on the other hand, one finds that the
amplitude of the Bragg peak is modified, and a background signal
appears which is due to light scattering by the condensate
fraction. This signal is the signature of the superfluid phase,
and it arises from the coherent effects of tunneling and
light-induced hopping. We also notice that in the superfluid phase
only the first diffraction order is visible. This is due to the
increased width of the atomic wavefunction, which yields a faster
decaying Debye-Waller factor $J_0(\V{q})$. We remark that higher
diffraction orders would be visible if the superfluid regime was
accessed by keeping the lattice depth constant, for instance by
ramping down the on-site energy using a Feshbach resonance. The
Bragg signal as a function of the lattice depth is reported in
Fig.~\ref{fig:Intensity:1}, showing the appearance of the
background signal as the superfluid regime is approached.

\section{Conclusions}\label{Sec:4}

We have discussed Bragg spectroscopy of ultracold bosonic atoms in
an optical lattice, focussing on the signatures of the
Mott-insulator and superfluid quantum state in the scattered
photons. A full quantum theory for the atoms and photons dynamics
and interactions has been developed, allowing us to identify the
various contributions to the detected signals. We have
characterized the Bragg scattering signal, for the parameters
sweeping across the transition from the Mott-insulator to the
superfluid quantum state. In particular, the contribution of
light-induced hopping, arising from atomic recoil due to photon
scattering, has been put into evidence. This term has
been neglected in previous theoretical
treatments~\cite{ReyBraggScattering,Mekhov_Maschler}. In this work
we have shown that its contribution can interfere with ordinary
tunneling between sites thereby significantly affecting the
spectroscopic signal. Its effect is visible in the behavior of the
height of the peaks in the spectrum as a function of the emission
angle, and it has been singled out by comparing the spectrum
evaluated when this effect is discarded. This effect can be
revealed experimentally in large systems, according to the
analytical theory we develop by extending the one derived
in~\cite{ReyBraggScattering,ReyPhd}, and in small systems, as we
observe by numerically evaluating the spectrum for a lattice of 7
atoms. It is interesting to consider whether such properties can
be used as resources for photonic interfaces based on
strongly-correlated atoms in optical lattices.

This analysis has been made in the linear regime, assuming a weak probe and far-off resonance both from the atoms and from the frequency of the lattice beam. Using instead Bragg beams at the same frequency as the optical lattice, wave-mixing effects are expected, as reported for instance  in~\cite{Grynberg,Phillips,Grynberg2}. In addition, optical lattices have been discussed in the literature as a possible realization of photonic bandgap materials~\cite{Deutsch,Lagendijk,Lambropoulos-Review,Pritchard,Artoni,Carusotto,Rist09,Antezza}. An interesting question is how such photonic properties are modified when the many-body quantum state of the atoms is relevant to the atom-photon interactions dynamics. When the light is close to resonance with the atoms, hence in the dissipative regime, the state of the atoms is significantly heated up. On the other hand, interesting photon-photon correlations could be observed, due to interference in multiple scattering by the atoms, see e.g. Ref.~\cite{Rist}.

We remark that, while monitoring the state of the gas by means of photons is attractive, on the other hand Bragg spectroscopy modifies the atomic system, as the recoil imparted by the scattered photon significantly perturbs the state of the atomic gas. It would be desirable to identify schemes, such as quantum-non-demolition type of measurements~\cite{QND,Polzik}, which can allow one to measure the relevant quantities in a non-invasive way. This may permit one to implement feedback mechanisms~\cite{Mekhov_QND,Eschner_Feedback}, which would allow one to  prepare other nonclassical states of the atomic gas.

\acknowledgements

The authors acknowledge Immanuel Bloch, Iacopo Carusotto, Igor Mekhov, Wolfgang Schleich, and Stefano Zippilli
for stimulating discussions and helpful comments. This work was supported by the European Commission (EMALI, MRTN-CT-2006-035369; Integrated Project SCALA, Contract No.\ 015714), by the European Science Foundation (EUROQUAM "CMMC"), and by the Spanish Ministerio de Educaci\'on y Ciencia (Consolider-Ingenio 2010 QOIT,  CSD2006-00019; QNLP, FIS2007-66944; Ramon-y-Cajal; Acci{\'o}n Integrada HU2007-0013. G.M. is supported by the DFG (German Research Council) with a Heisenberg professorship. C.M. thanks ICFO - The Institut for Photonic Sciences in Barcelona for hospitality in the period when this work was started.


\begin{thebibliography}{99}
\bibitem{Footnote} Equation~(\ref{cross:section}) is reported with
the density of states at the value of the laser frequency
$\omega_L$. This approximation is justified, as the range of
frequencies of the final states is of the order of the recoil
frequency (from few to tens KHz), while the laser frequency is in
the optical regime. The same approximation will be applied to the
vacuum coupling strengths $C_\lambda$, Eq.~(\ref{C:mu}) entering the
interaction Hamiltonian $H_{\rm int}'$, Eq.~(\ref{eqn:Heff}).
\bibitem{Footnote:Linewidth}
The linewidth of the excitations is essentially determined by the spontaneous decay rate of the excited state. In the present treatment we are assuming $\gamma^{\prime} T\ll 1$, where $\gamma^{\prime}$ is the rate of incoherent scattering and is of the order $\gamma^{\prime}\sim \gamma |C_L\alpha|^2/\Delta^2$.
\end{thebibliography}
\end{document}